# Beam data modeling of linear accelerators (linacs) through machine learning and its potential applications in fast and robust linac commissioning and quality assurance

Wei Zhao,[1] Ishan Patil,[1] Bin Han,[1] Yong Yang,[1] Lei Xing,[1]* Emil Schüler,[1,2]*

[1]Stanford University, Department of Radiation Oncology, Stanford, CA 94305 USA.

[2]The University of Texas MD Anderson Cancer Center, Department of Radiation Physics, Houston, TX 77030, USA

Email address:
Wei Zhao: zhaow85@stanford.edu
Ishan Patil: ishanpatil1994@gmail.com
Bin Han: hanbin@stanford.edu
Yong Yang: yongy66@stanford.edu
Lei Xing: lei@stanford.edu
Emil Schüler: eschueler@mdanderson.org

Correspondence to: Professor Lei Xing, Ph.D., DABR. Department of Radiation Oncology, Stanford University, 875 Blake Wilbur Drive Room G233, Stanford, CA 94305-5847, Ph: (650) 498-7896. Email: lei@stanford.edu; Dr Emil Schüler, Ph.D., Department of Radiation Physics, Radiation Oncology Division, UT MD Anderson Cancer Center. Email: eschueler@mdanderson.org

The authors would like to thank the Division of Medical Physics at University of California, San Francisco, the Division of Medical Physics at University of California, San Diego, the Division of Medical Physics and Engineering at University of Texas, Southwestern, BC Cancer – Victoria, University of Victoria, and the department of Radiation Oncology at Stanford University for their contribution of beam data. This work was partially supported by NIH/NCI (1R01CA176553, 1R01CA223667 and 1R01CA227713) and a Faculty Research Award from Google Inc.

## Abstract

**Background and purpose:** To propose a novel machine learning-based method for reliable and accurate modeling of linac beam data applicable to the processes of linac commissioning and QA.

**Materials and methods:** We hypothesize that the beam data is a function of inherent linac features and percentage depth doses (PDDs) and profiles of different field sizes are correlated with each other. The correlation is formulated as a multivariable regression problem using a machine learning framework. Varian TrueBeam beam data sets (n=43) acquired from multiple institutions were used to evaluate the framework. The data sets included PDDs and profiles across different energies and field sizes. A multivariate regression model was trained for prediction of beam specific PDDs and profiles of different field sizes using a 10x10cm$^2$ field as input.

**Results:** Predictions of PDDs were achieved with a mean absolute percent relative error (%RE) of 0.19–0.35% across the different beam energies investigated. The maximum mean absolute %RE was 0.93%. For profile prediction, the mean absolute %RE was 0.66-0.93% with a maximum absolute %RE of 3.76%. The largest uncertainties in the PDD and profile predictions were found at the build-up region and at the field penumbra, respectively. The prediction accuracy increased with the number of training sets up to around 20 training sets.

**Conclusions:** Through this novel machine learning-based method we have shown accurate and reproducible generation of beam data for linac commissioning for routine radiation therapy. This method has the potential to simplify the linac commissioning procedure, save time and manpower while increasing the accuracy of the commissioning process.

# Introduction

Radiation therapy is a central treatment strategy in the management of cancer and is indicated in more than 50% of all cancer cases [1-4]. The majority of the patients who undergo radiation therapy will do so through external photon radiation therapy utilizing linear accelerators incorporated into a rotating gantry for highly conformal dose deposition within the target [1]. The introduction of these clinical linear accelerators (linacs) into a radiation oncology clinic for cancer treatment will be preceded by rigorous testing and characterization related to safety, mechanical operation, and dosimetry [5-11]. The data acquired during this commissioning process will then be used throughout the lifetime of the machine for verification of proper linac operation, in the continuing quality assurance (QA) process, and as input to the treatment planning software (TPS) used for patient dose calculations[5, 12]. Due to the importance of the commissioning process, the American Association for Physicists in Medicine (AAPM) has provided guidelines and recommendations on equipment and procedures for the acquisition of beam data for the dosimetric commissioning of linacs [5]. However, no method for verifying acquired data quality is given.

The method of choice for data verification used clinically is instead by comparing the acquired data to vendor-supplied data sets [5, 13-15]. These data sets are usually comprised of data that have been averaged across a handful of machines of the same model. The concept of using the vendor-supplied data sets relies on the notion of linacs being the mirror of each other. However, no two linacs are identical due to e.g. inherent uncertainties in the manufacturing of the complex systems that are part of the overall linac design [16]. It is therefore not advisable to use these data sets directly in the TPS for patient dose calculation, nor is it advisable to substitute commissioning data sets with vendor-supplied data sets [16].

Despite the risk of introducing uncertainties in the beam calculations by using averaged data from other linacs, some institutions use the vendor-supplied data sets directly in their TPS for patient dose calculations and use the commissioning process only to verify that the measured beam parameters are within a certain institutional tolerance from these data sets [5, 17, 18]. The rationale for this approach is that radiation therapy is a complex process with multiple potential sources of errors. Specifically, the commissioning process is considered one of the most complex processes in radiation oncology today and with significant potential for the introduction of errors [5, 19-25]. National surveys have shown that a large fraction of institutions fail to deliver the prescribed dose within clinically acceptable tolerance limits and pretreatment QA for identification of unacceptable plans is very important [26]. Many of these failures have been associated with errors in the commissioning data [27]. By using the vendor-supplied beam data, the institutions therefore minimize the risk of human errors in the data handling while accepting the error introduced by the assumption that all linacs from the same vendor and model are identical.

While no two linacs are identical, the parameters measured during the linac commissioning procedure are highly correlated and they are a function of many inherent linac features. For example, the percentage depth dose (PDD) is a function of beam energy, field size, source-to-surface distance (SSD), beam filtering, etc. By measuring a specific PDD, it would therefore be possible to predict a PDD with a different set of features as this correlation can be modeled using a mathematical model based on sample data, known as "training data". Once the correlation is modeled, it can be used to make predictions without being explicitly formulated to obtain PDDs. Developing machine learning models for use in the commission procedure would provide needed assurance and reduce time-consuming and repetitive tasks that are inherent to the commissioning process [28]. Moreover, as highlighted in AAPM's TG-100 report on radiation therapy quality management, increasing the utilization of automation and computerization in the clinic would reduce or even eliminate many of the error-prone tasks that are present in clinical medical physics [29]. Moving the commissioning process into a more automated approach would also help to address one of the issues related to the critical shortage of linacs in many low to middle-income countries where an easier commissioning process would help address the operational and staffing costs required to setup and maintain the linac [30, 31].

The goal of this study was therefore to establish a novel data-driven strategy using machine learning for prediction of clinical linear accelerator beam data for measurement verification and data generation during the commissioning and calibration processes of linacs in routine clinical operation.

## Methods and Materials

### Problem formulation

The parameters included in the linac beam data are highly correlated and they are a function of many inherent linac features (multivariable functions). For example, the PDD can be affected by many variables and need to be measured at all possible configurations (such as different beam energies, field sizes, etc.) which involve numerous repeated measurements with slightly updated linac features. These PDDs are highly correlated intrinsically and it is therefore possible to infer a PDD for a specific linac feature using

other measured PDDs. The problem can be formulated as follows (detailed in Appendix A):

$$Y = X\beta + E. \tag{1}$$

Where matrix $Y = (y_{ij}: i = 1, ..., N; j = 1, ..., M)$ and $y_{ij}$ represents the $j$th depth dose for the $i$th PDD curve that has $M$ depth dose measurements. Matrix $X = (x_{ik}: i = 1, ..., N; k = 1, ..., K)$ and $x_{ik}$ represents the $k$th measurement point for the $i$th PDD sample which has the same energy but different field sizes. Matrix $\beta$ is the weight that to be determined and $E$ is the random noise during the measurement.

With $N$ pairs of depth dose measurement samples and the corresponding PDD full curves, one can solve the above equation to obtain the coefficients matrix $\beta$, and with the predetermined $\beta$ and a new depth dose measurement, we can infer the corresponding PDD full curve at a different setting. Note that $\beta$ is determined using measurement samples and the corresponding PDD curves, both of which are directly associated with the specific linac features. Hence $\beta$ is related to these features and should be uniquely determined by these features.

**Machine learning-based linac beam data prediction**

To solve (1), one can minimize the residual summation of squares between the observed responses $(Y' = Y - E)$ in the training datasets, and the responses predicted by the approximation $(\hat{Y} = X\beta)$. Mathematically it solves a problem of the form:

$$\beta = arg\min_{\beta} \|\hat{Y} - Y'\|_2^2, \tag{2}$$

i.e.,

$$\beta = arg\min_{\beta} \|X\beta - Y'\|_2^2. \tag{3}$$

Before linac commissioning, vendors have usually performed careful calibrations with respect to vendor-supplied averaged data from the same linac model previously installed elsewhere. Hence, it is possible that columns in the sampling matrix $X$ have an approximate dependence and may become close to a singular. Meanwhile, the measured commissioning data $Y'$ have inherent random noise $E$, and as a result, the least square optimization in (3) may become sensitive to the random noise, resulting in a large variance. To mitigate this problem, we further introduce a penalty term to the least square data-fidelity term. The penalty term is a $l_2$-norm which is applied to the coefficients matrix $\boldsymbol{\beta}$ to ensure its Euclidean distance to be optimized together with the data-fidelity term. In this case, the optimization problem is formulated as:

$$\boldsymbol{\beta} = arg\min_{\boldsymbol{\beta}} \|X\boldsymbol{\beta} - Y'\|_2^2 + \alpha\|\boldsymbol{\beta}\|_2^2 \qquad (4)$$

$\alpha$ is a complexity parameter that balances the data-fidelity term and the penalty term. A higher $\alpha$ suggests a stronger regularization on the matrix $\boldsymbol{\beta}$. In this work an $\alpha$ of 0.1 was used.

In this study, based on the previous linac commissioning and annual QA datasets, we performed supervised learning for predictions of beam data. For each of the linac features, a unique matrix $\boldsymbol{\beta}$ was obtained using (4) and then applied to the new linac.

**Datasets**

Beam data acquired during water tank measurements for commissioning/annual QA (n=43) were collected from Varian TrueBeams from 9 institutions. The IBA Blue Phantom or the IBA Blue Phantom 2 was used for data collection (IBA Dosimetry, Germany). The data sets included PDDs and profiles across different energies, field sizes, and depths. Beams with flattening filter (WFF) and flattening filter free (FFF) were included in the database. All scanning data were acquired with a SSD of 100 cm. The choice of detector varied between institutions and both cylindrical ion chambers (up to 6 mm diameter, 0.13 cm$^3$ active volume) and diode detectors (Sun Nuclear EDGE) were used. For cylindrical chambers, pre-processing of the PDD data sets included shifting of the data by 0.6 times the radius of chamber ( $0.6r_{cav}$ ) used as per TG-51 recommendations [21]. Normalization of the PDD datasets was performed at 10 cm depth to eliminate the effect of the noise around the depth of maximum dose ($d_{max}$). For visualization purposes in the results section, the data presented are re-normalized to $d_{max}$. The profile data sets were centered and normalized to the central axis.

**Implementation details**

After data pre-processing, the data sets were separated by energy, field size, and type (PDD or profile). Only profiles collected at 10 cm depth were included into the prediction model. The Scikit-learn machine learning toolkit [32] was employed for model training with ridge regularization with $\alpha = 0.1$. Prediction models were trained using the full set of data for a specific situation (energy, field size, and type) but leaving one data set out for validation (Fig. 1). This leave-one-out approach was repeated until all data sets had been used for validation purposes. The predictions were evaluated by calculating the percent relative error ( $\%RE = 100\% * \frac{X_{predicted} - X_{measured}}{X_{measured}}$ ) along the PDD or profile.

To evaluate the performance of the prediction models as compared to utilizing the vendors supplied averaged data sets, the %RE was calculated for each individual data set included in the training of the models, such that

$$\%RE = 100\% * \frac{X_{vendor} - X_{measured}}{X_{measured}}$$

In addition, distance-to-agreement (DTA) and local gamma analysis using criteria of 1%/1mm and 2%/1mm [17, 33] were used to evaluate the accuracy of the PDDs and dose profiles predicted by the proposed algorithm at different beam settings. The gamma passing (GP) rate was calculated for the predicted PDDs and profiles at different field sizes and beam delivery settings.

**Dependence of number of datasets used in prediction**

To investigate the dependence on the number of training sets included during the model training and the accuracy of the prediction, a modified model training approach was utilized. During the training of the model, the number of training sets were varied from 1 to the full data set minus the validation set. For each validation set, multiple models were generated for every combination of the training sets so as to remove any bias from the included data sets. Predictions were performed for each model and the data were averaged before the next data set was extracted to be used as the validation set. This was continued until every combination had been tested and the number of data sets included in the model had been varied from 1 to the full data set minus 1.

## Results

Through the methodology described above, predictions of PDDs and profiles for different combinations of field sizes and depths were performed. Figure 2 shows an example of the prediction of beam data for 6 MV WFF, 6 MV FFF, and 10 MV FFF beam for 4x4 cm$^2$ and 30x30 cm$^2$ fields utilizing the energy-specific data of the 10x10 cm$^2$ field as model input.

**Dependence of the number of datasets used in prediction**

The number of data sets required for accurate predictions were evaluated using the most abundant data set combinations in our database which were PDDs of 4x4 cm$^2$, 10x10 cm$^2$, and 30x30 cm$^2$, and profiles of 10x10 cm$^2$ and 30x30 cm$^2$. Model training was performed with one to the full range of data sets minus 1 in our database as training data (see Methods section). Figure 3 shows the mean absolute %RE for the different beam energies with 10x10 cm$^2$ data as model input. The mean absolute and the range of absolute %RE decreases with an increasing number of data sets included in the model training for both PDD (Fig. 3 A and B) and profile prediction (Fig. 3C). Based on 30x30 cm$^2$ PDD and profile predictions, a plateau was reached at around 20 data sets.

**Model evaluation**

The mean absolute %RE across the entire PDD ranged between 0.19% and 0.35% for PDD predictions using 10x10 cm$^2$ field size as model input, with a maximum absolute %RE ranging between 0.29% and 0.93% (Tab. 1). Beams with flattening filters showed higher prediction accuracy than FFF beams, and PDD predictions for 30x30 cm$^2$ showed higher accuracy than 4x4 cm$^2$ field size. This latter is most likely due to the fewer number of data sets available for 4x4 cm$^2$ fields. Since there were at least 18 testing samples for different beam settings, the mean absolute %RE across the entire profile for each testing sample at the 30x30 cm$^2$ field ranged between 0.66% and 0.93% with a maximum absolute %RE between 1.6% and 3.76%. No dependence on beam filtering was found for profile prediction.

The PDD prediction accuracy was minimally affected by the direction of prediction for a given field size pair (Tab. 2). However, for profile prediction, larger importance for the input/output data combination was found (Tab. 3). Using the profile of a 10x10 cm$^2$ field for profile prediction of a 30x30 cm$^2$ field resulted in a lower absolute %RE than if using the larger field size for prediction of the smaller field size. However, when evaluating the mean absolute %RE across the central 80% of the field, no dependence on the direction of prediction was found. Quantitative evaluation of dose distribution using gamma analysis for both the predicted PDDs of 4x4 cm$^2$ and 30x30 cm$^2$ fields at different beam delivery setting yield clinically acceptable GP rates using criteria of 2%/1mm and 1%/1mm (Tab. A.1). Analysis of the uncertainty along the predicted PDDs and profiles, and the sensitivity of the model using %RE and DTA are detailed in Appendix B.

To evaluate the performance of the prediction models as compared to utilizing the vendors supplied averaged data sets, the %RE was calculated between each individual data set included in the training of the models and the vendor-supplied data sets. For all energies and field sizes investigated, an increased %RE was found if using the vendor-supplied data sets as compared to the predicted beam data. The mean %RE for the entire PDDs for all energies and field sizes increased from 0.25% to 0.42% and the maximum error increased from 0.58% to 1.5%. The mean %RE for the entire profiles for all energies increased from 0.79% to 1.48% and the maximum error increased from 2.21% to 3.9%.

## Discussion

Utilizing the methodology described in this paper we have shown accurate predictions of PDDs and profiles for all beam types investigated. With 10x10 cm$^2$ beam data as model input, predictions of PDDs for field sizes of 4×4 cm$^2$ and 30×30 cm$^2$ were achieved with a %RE of < 1% for all energies, with similar accuracy found for profile predictions. In practice, the tolerances on agreement between PDD and profile generated by the TPS and measured ones are considered to be 1% for the local dose and within 1mm for the DTA, whichever is applicable [17]. Based on the gamma analysis and composite analysis using DTA and dose difference, we found the differences between dose distributions predicted by the proposed algorithm and measured ones are within the tolerances for almost all sampling points, except the build-up region which usually has larger tolerance [17]. Indeed, the largest discrepancy in the predictions was found in the build-up region of the PDDs and at the penumbra of the profiles. The discrepancy in these regions is most likely related to i) positional uncertainties of the detector and ii) differential volume averaging across the data sets used for model training. Concerning i), due to the rapidly changing dose per unit distance in these regions, any positional error of the detector will cause a large deviation in measured dose. This in turn will introduce a larger uncertainty in these regions when combining multiple data sets as seen here. Concerning ii), the choice of detector between the different institutions differed and the chamber type and size ranged from cylindrical ionization chambers with an active volume of 0.13 cm$^3$ to diode detectors with an active volume of 0.019 mm$^3$. A larger chamber is more susceptible to volume averaging, especially at the aforementioned regions [34]. Furthermore, volume averaging will have a higher impact with FFF beams as compared to flattened beams, which could be the reason for the larger uncertainty in the prediction of these beams. The effect of volume averaging could potentially be mitigated by incorporating the chamber size as a feature in the prediction model. However, the number of data sets did not allow this kind of distinction.

Around 20 data sets were needed to stabilize the predictions. However, the number of data sets included reaching a certain level of accuracy will be dependent on the variability and uncertainty in the training data. More data sets would further strengthen the predictions as they would allow for inclusion of more scenarios into the model. While the use of real commissioned data adds strength to the prediction, it is difficult to incorporate the many different scenarios and combinations of parameters that affect the beam data. Furthermore, because of the nature of the data, all the data was within tolerance concerning e.g. flatness and symmetry [12]. Having beam data that is within tolerance with a nominal field size but when measured at a smaller/larger field sizes the beam data was out of tolerance would be valuable to have included into the model. A properly trained model would most likely be able to pick up on features of the measured data that could predict an out-of-tolerance at a different field size. However, such data is most often not saved after beam tuning and therefore we did not have access to data that was outside of tolerance. Monte Carlo (MC) generated data could be a powerful tool to employ to allow sampling of scenarios where tolerance was breached. Data sets with more tightly sampled data could be generated with controlled variations such as misalignments of jaw position, chamber position, chamber angle, collimator angle, gantry angle, SSD, etc., which could be used in the model training. Through such an approach the uncertainties in the model introduced through human error and setup uncertainties could be excluded and therefore making the predictions more powerful and robust.

When testing the model for different scenarios of erroneous input data, input data taken at the wrong SSD had a large effect on the predictions. With properly set thresholds and action limits, this error would have been easily detected by an automated system. A depth offset of the input data had a smaller effect and would be within the uncertainties of the overall predictions, potentially even in the build-up region. Detecting this error as a true error would require more stringent thresholds to warn the operator of potential error. However, we envision that the inclusion of more data sets both from other institutions and from data sets generated through MC will reduce the uncertainties of the predictions and would therefore allow for easier detection of input data error. An interesting aspect in the sensitivity analysis was the robustness of the model for truncated data sets. The prediction accuracy did not change compared to using the full data set even when the input data was truncated at 50 mm depth. This opens the question about which features of the PDD are important and also with what spatial resolution is needed to sample the PDD in order to make accurate predictions? While these questions are beyond the scope of the current study, they will be important to answer in the optimization procedure of the model and in the development of a prediction and verification system.

In this study, we chose to only include measured data acquired during commissioning and annual QA of Varian TrueBeams from multiple institutions. One limitation of using data from multiple institutions is the variation of the collected data in terms of field sizes, energies, and depths. Most institutions would collect a subset of the full range of data combinations in order to save time and reduce the invested effort. Due to the lack of proper guidelines in this data collection, different institutions have different strategies which have developed as a result of the history of the institution. For this reason, we were only able to build models for predictions of PDDs for combinations of 4x4 cm$^2$, 10x10 cm$^2$, and 30x30 cm$^2$ fields and for profiles for combinations of 10x10 cm$^2$ and 30x30 cm$^2$ at 10 cm depth. The main predictions and evaluations were performed with input data from a 10x10 cm$^2$ field. This setup was chosen as it represents the setup recommended in the AAPM TG-51 report for reference dosimetry (SSD=100 cm, field size of 10x10 cm$^2$ at surface)[21]. However, the model is not limited to this arrangement and similar accuracy is achieved independent of how the input and output data is combined.

A practical scenario of using the proposed method would be at the annual QA for a specific linac model (e.g. TrueBeam) [35-37]. With the pretrained model using routinely obtained beam data at two specific settings (such as 10×10 cm$^2$/30×30 cm$^2$), one only need to measure the dose distribution of the machine with 10×10 cm$^2$ field size and the dose distribution at the same energy level but different field sizes can then be predicted using the model during the annual QA procedure. In this study, since the testing samples including both annual QA and commissioning data, the method can also be applied to linac commissioning. However, if the correlation of the testing dose distribution between two field sizes is very different from that of the beam data used for model training (e.g. in case the hardware of a linac model gets updated), the predictive model may yield inferior results. More importantly, the algorithm should be considered as a medical device from a legal perspective and should be subjected to pre-market review and post-market surveillance, where the premarket review requires to demonstrate effectiveness in practice and the post-market surveillance requires lifetime monitoring and risk management.

The choice of only using Varian TrueBeams in this study was due to the greater availability of scanned water tank data with these linacs compared to other currently used linacs, and not due to limitations of the methodology itself. Varian TrueBeams also represents less inter-machine variability in beam data as compared to other radiotherapy machines which allow for a lower number of training data sets needed for proper model training [6, 18, 38]. Due to the low inter-machine variability in beam data, many institutions therefore opt for using vendor-supplied beam data in their treatment planning system [38, 39]. The use of the vendor-supplied data sets which usually encompass averaged data across a handful of linacs of the same configurations, relies on the assumption of linacs being the mirror of each other. However, inherent uncertainties are inevitable in the manufacturing of the complex linac systems and no two linacs are identical [16]. It is therefore not advisable to use these data sets directly in the TPS for patient dose calculation, nor is it advisable to substitute commissioning data sets with vendor supplied data sets [16]. Utilizing instead the model presented in this study would not only allow a faster commissioning and QA procedure, but would also decrease the error in the machine beam data as was shown in Table 1.

In conclusion, the work presented here promises to provide a data-driven scheme for generation of high accuracy beam data. The method has the potential to serve as a verification tool of acquired beam data, an error prediction tool for assessing how variations of beam data for one field size affects another, and finally as a way of reducing the numerous measurements that are needed for proper implementation of a new linac into the clinical workflow without compromising the quality of the data. The successful implementation of the methodology would greatly simplify the linac commissioning procedure, save time and manpower while increasing the accuracy of the commissioning and annual QA process.

# Reference

1. Miller, K.D., et al., *Cancer treatment and survivorship statistics, 2016.* CA: a cancer journal for clinicians, 2016. **66**(4): p. 271-289.
2. Bentzen, S.M., *Preventing or reducing late side effects of radiation therapy: radiobiology meets molecular pathology.* Nature Reviews Cancer, 2006. **6**(9): p. 702-713.
3. De Ruysscher, D., et al., *Radiotherapy toxicity.* Nature Reviews Disease Primers, 2019. **5**(1): p. 1-20.
4. Yap, M.L., et al., *Global access to radiotherapy services: have we made progress during the past decade?* Journal of global oncology, 2016. **2**(4): p. 207-215.
5. Das, I.J., et al., *Accelerator beam data commissioning equipment and procedures: report of the TG‐106 of the Therapy Physics Committee of the AAPM.* Medical physics, 2008. **35**(9): p. 4186-4215.
6. Glide‐Hurst, C., et al., *Commissioning of the Varian TrueBeam linear accelerator: a multi‐institutional study.* Medical physics, 2013. **40**(3): p. 031719.
7. Farr, J., et al., *Development, commissioning, and evaluation of a new intensity modulated minibeam proton therapy system.* Medical physics, 2018. **45**(9): p. 4227-4237.
8. Netherton, T., et al., *Experience in commissioning the halcyon linac.* Medical physics, 2019.
9. Ciocca, M., et al., *Design and commissioning of the non‐dedicated scanning proton beamline for ocular treatment at the synchrotron‐based CNAO facility.* Medical physics, 2019. **46**(4): p. 1852-1862.
10. Liu, P.Z.Y., et al., *Development and commissioning of a full‐size prototype fixed‐beam radiotherapy system with horizontal patient rotation.* Medical physics, 2019. **46**(3): p. 1331-1340.
11. Teo, P.T., et al., *Application of TG‐100 risk analysis methods to the acceptance testing and commissioning process of a Halcyon linear accelerator.* Medical physics, 2019. **46**(3): p. 1341-1354.
12. Klein, E.E., et al., *Task Group 142 report: Quality assurance of medical acceleratorsa.* Medical physics, 2009. **36**(9Part1): p. 4197-4212.
13. LaRiviere, P.D., *The quality of high-energy X-ray beams.* The British journal of radiology, 1989. **62**(737): p. 473-481.
14. Kosunen, A. and D. Rogers, *Beam quality specification for photon beam dosimetry.* Medical physics, 1993. **20**(4): p. 1181-1188.
15. Kalach, N. and D. Rogers, *Which accelerator photon beams are "clinic‐like" for reference dosimetry purposes?* Medical physics, 2003. **30**(7): p. 1546-1555.
16. Das, I.J., C.F. Njeh, and C.G. Orton, *Vendor provided machine data should never be used as a substitute for fully commissioning a linear accelerator.* Medical physics, 2012. **39**(2): p. 569-572.
17. Venselaar, J., H. Welleweerd, and B. Mijnheer, *Tolerances for the accuracy of photon beam dose calculations of treatment planning systems.* Radiotherapy and oncology, 2001. **60**(2): p. 191-201.
18. Beyer, G.P., *Commissioning measurements for photon beam data on three TrueBeam linear accelerators, and comparison with Trilogy and Clinac 2100 linear accelerators.* Journal of applied clinical medical physics, 2013. **14**(1): p. 273-288.

19. Njeh, C., *Tumor delineation: The weakest link in the search for accuracy in radiotherapy.* Journal of medical physics/Association of Medical Physicists of India, 2008. **33**(4): p. 136.
20. Adnani, N., *Design and clinical implementation of a TG‑106 compliant linear accelerator data management system and MU calculator.* Journal of applied clinical medical physics, 2010. **11**(3): p. 12-25.
21. Almond, P.R., et al., *AAPM's TG‑51 protocol for clinical reference dosimetry of high‑energy photon and electron beams.* Medical physics, 1999. **26**(9): p. 1847-1870.
22. Fogliata, A., et al., *Flattening filter free beam from Halcyon linac: Evaluation of the profile parameters for quality assurance.* Medical physics, 2020.
23. Al Mashud, M.A., et al., *Photon beam commissioning of an Elekta Synergy linear accelerator.* Polish Journal of Medical Physics and Engineering, 2017. **23**(4): p. 115-119.
24. Narayanasamy, G., et al., *Commissioning an Elekta Versa HD linear accelerator.* Journal of applied clinical medical physics, 2016. **17**(1): p. 179-191.
25. Netherton, T., et al., *Experience in commissioning the halcyon linac.* Medical physics, 2019. **46**(10): p. 4304-4313.
26. Kry, S.F., et al., *Independent recalculation outperforms traditional measurement‑based IMRT QA methods in detecting unacceptable plans.* Medical physics, 2019. **46**(8): p. 3700-3708.
27. Ibbott, G.S., et al., *Challenges in credentialing institutions and participants in advanced technology multi-institutional clinical trials.* International Journal of Radiation Oncology* Biology* Physics, 2008. **71**(1): p. S71-S75.
28. Amols, H.I., F. Van den Heuvel, and C.G. Orton, *Radiotherapy physicists have become glorified technicians rather than clinical scientists.* Medical Physics-New York-Institute of Physics, 2010. **37**(4): p. 1379.
29. Huq, M.S., et al., *The report of Task Group 100 of the AAPM: Application of risk analysis methods to radiation therapy quality management.* Medical physics, 2016. **43**(7): p. 4209-4262.
30. Atun, R., et al., *Expanding global access to radiotherapy.* The lancet oncology, 2015. **16**(10): p. 1153-1186.
31. Van Dyk, J., E. Zubizarreta, and Y. Lievens, *Cost evaluation to optimise radiation therapy implementation in different income settings: A time-driven activity-based analysis.* Radiotherapy and Oncology, 2017. **125**(2): p. 178-185.
32. Pedregosa, F., et al., *Scikit-learn: Machine learning in Python.* Journal of machine learning research, 2011. **12**(Oct): p. 2825-2830.
33. Low, D.A., et al., *A technique for the quantitative evaluation of dose distributions.* Medical physics, 1998. **25**(5): p. 656-661.
34. Bouchard, H. and J. Seuntjens, *Ionization chamber‑based reference dosimetry of intensity modulated radiation beams.* Medical physics, 2004. **31**(9): p. 2454-2465.
35. Chan, M.F., et al., *Visual analysis of the daily QA results of photon and electron beams of a trilogy linac over a five-year period.* International journal of medical physics, clinical engineering and radiation oncology, 2015. **4**(4): p. 290.
36. Li, Q. and M.F. Chan, *Predictive time-series modeling using artificial neural networks for Linac beam symmetry: an empirical study.* Annals of the New York Academy of Sciences, 2017. **1387**(1): p. 84.
37. Chan, M.F., A. Witztum, and G. Valdes, *Integration of AI and Machine Learning in Radiotherapy QA.* Frontiers in Artificial Intelligence, 2020. **3**: p. 76.

38. Tanaka, Y., et al., *Do the representative beam data for TrueBeam™ linear accelerators represent average data?* Journal of applied clinical medical physics, 2019. **20**(2): p. 51-62.
39. Chang, Z., et al., *Commissioning and dosimetric characteristics of TrueBeam system: composite data of three TrueBeam machines.* Medical physics, 2012. **39**(11): p. 6981-7018.

# Appendix A

For a specific beam energy, we hypothesize that the PDD curve at a specific field size can be predicted by using a set of measured PDDs at a different field size, e.g. predicting PDD$_{30\times30}$ at field size 30×30 cm² with measurements from PDD$_{10\times10}$ at field size 10×10 cm². Namely, PDD$_{30\times30}$ is mapped to PDD$_{10\times10}$ with an unknow function f, which can be denoted as:

$$PDD_{30\times30} = f(PDD_{10\times10}). \tag{A.1}$$

This unknown function should be attributed to many of the inherent linac features and may not have an explicit analytical form. However, given commissioning datasets from a set of linacs, it is possible to find an empirical mapping from the above function. To this end, we denote the to-be predicted PDD distribution as $\mathbf{y} \in R^M$, with M the number of measurement points of the depth dose, and the sampling measurement of an input depth dose at a different setting as $\mathbf{x} \in R^K$, with $K$ as the number of sampling point of the specific setting. Note that $K$ and $M$ do not need to be the same and the measurement locations do not need to be consistent either. Suppose we have a set of $N$ input depth dose samples $\{\mathbf{x_1}, \mathbf{x_2}, \ldots, \mathbf{x_N}\} \subset R^K$ at the specific setting, with the corresponding PDD response data sets $\{\mathbf{y_1}, \mathbf{y_2}, \cdots \mathbf{y_N}\} \subset R^M$. Assume a linear model $\mathbf{y}_i = (y_{ij}: j = 1, \ldots, M)$, $\mathbf{x}_j = (x_{jk}: k = 1, \ldots, K)$:

$$y_{ij} = x_{i1}\beta_{1j} + x_{i2}\beta_{2j} + \cdots + x_{iK}\beta_{Kj} + \epsilon_{ij} = \sum_{k=1}^{K} x_{ik}\beta_{kj} + \epsilon_{ij}, \tag{A.2}$$

Here $y_{ij}$ is the *j*th depth dose for the *i*th PDD curve that has M depth dose measurements. $x_{ik}$ is the *k*th measurement point for the *i*th PDD sample which has the same energy but different field sizes. $\beta_{kj}$ is the weight that to be determined. $\epsilon_{ij}$ is the random noise during the measurement.

The above model suggests each depth dose in the full PDD curve of the N datasets can be expressed as a weighted summation of the *K* measurement points with a different setting. Based on the model, the PDD curves in the full datasets can be written as:

$$\begin{pmatrix} y_{11} & \cdots & y_{1M} \\ \vdots & \ddots & \vdots \\ y_{N1} & \cdots & y_{NM} \end{pmatrix} = \begin{pmatrix} x_{11} & \cdots & x_{1K} \\ \vdots & \ddots & \vdots \\ x_{N1} & \cdots & x_{NK} \end{pmatrix} \begin{pmatrix} \beta_{11} & \cdots & \beta_{1M} \\ \vdots & \ddots & \vdots \\ \beta_{K1} & \cdots & \beta_{KM} \end{pmatrix} + \begin{pmatrix} \epsilon_{11} & \cdots & \epsilon_{1M} \\ \vdots & \ddots & \vdots \\ \epsilon_{N1} & \cdots & \epsilon_{NM} \end{pmatrix}. \quad (A.3)$$

In a matrix formulation, we have,

$$\boldsymbol{Y} = \boldsymbol{X}\boldsymbol{\beta} + \boldsymbol{E}. \quad (A.4)$$

# Appendix B

The uncertainty along the PDDs was greatest in the build-up region for all predictions (Fig. A.1 and A.2). For PDD predictions of 4x4 cm$^2$ fields, the dose differences of most sampling points are smaller than 1% among the five beam types investigated, with few points having larger uncertainty in prediction with depth. For PDD predictions of 30x30 cm$^2$ fields, 6 MV and 6 MV FFF beams showed largest variability. DTA for the PDD at different beam delivery settings between the predictions and measurements were calculated for 4x4 cm$^2$ (Fig. A.3) and 30x30 cm$^2$ fields (Fig. A.4). Composite analysis using a pass-fail criterion of both the DTA and dose differences (Fig. A.1 and A.2) indicates most points pass a tolerance of 1%/1mm, except few points in the build-up region. For the profile predictions, the largest uncertainty was found at the penumbra (Fig. A.5). FFF beams had lower prediction accuracy compared to flat beams.

The sensitivity of the model on incorrect data input was tested on three different cases with known intentionally induced errors: Input data acquired with SSD of 90 cm for model trained on data acquired with SSD of 100 cm; Input data with a 2mm depth offset; Input data truncated at 50 mm depth. For these tests, the model was trained using the 6MV flat beam data with 10x10 cm$^2$ PDDs at 100 cm SSD as input data and 30x30 cm$^2$ PDDs at

100 cm SSD as output data. The average absolute %RE was evaluated as well as the %RE at 10 cm depth.

As shown in Fig. A.6, using input data with SSD = 90 cm introduced an error of the prediction that was dependent on depth with a mean %RE of 1.05% and %RE at 10 cm depth of 1.89%. Using input data with a depth offset of 2mm introduced small errors in the overall predictions (<1%) with maximum error in the buildup region. Truncated input data (truncated at 50 mm depth) had no effect on the predicted PDD of a 30x30 cm$^2$ field.

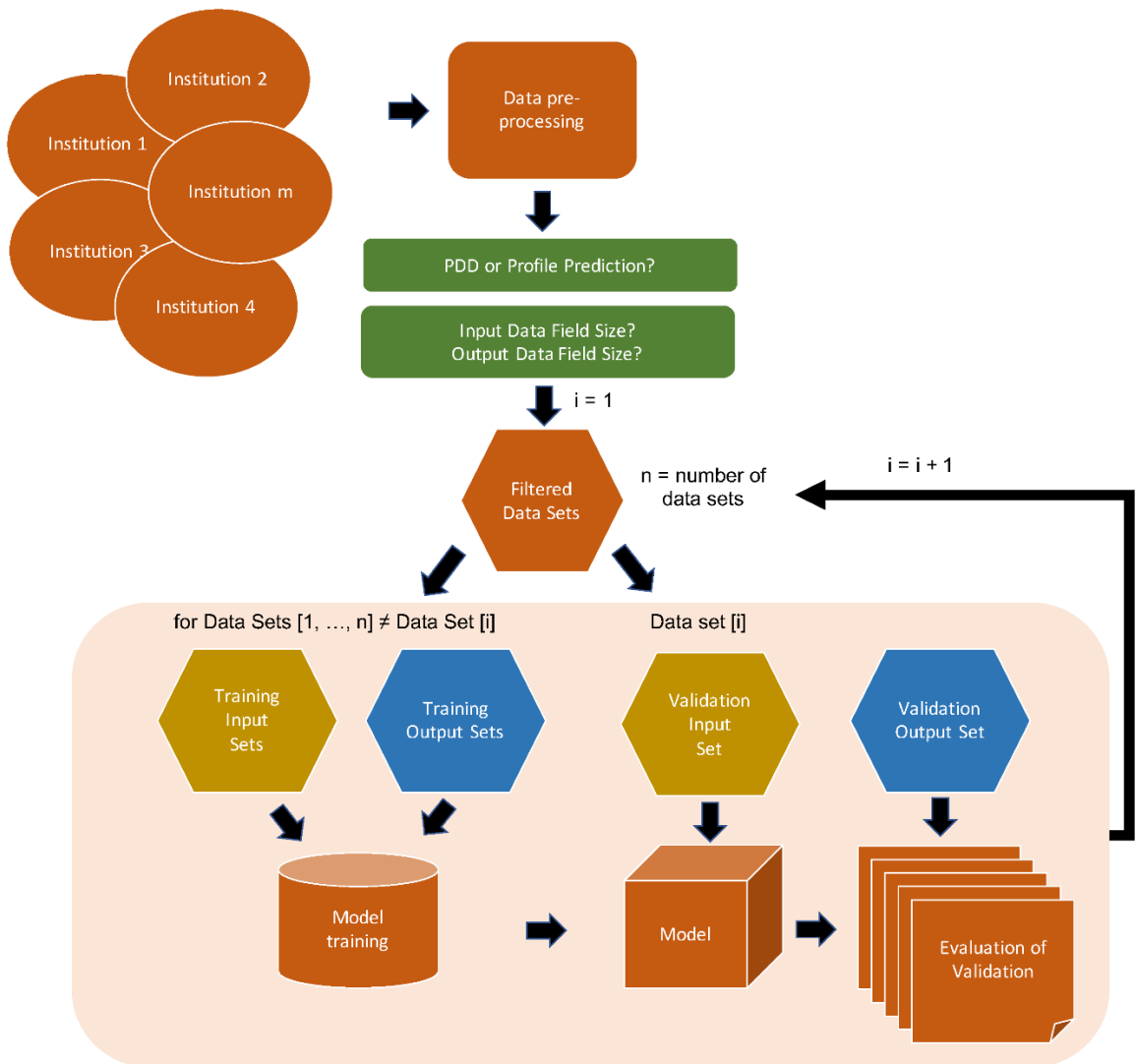

Figure 1. Workflow for model training and prediction. Beam data was collected from different institutions and pre-processed accordingly. The data was filtered according to the chosen prediction type (PDD or profile prediction), input field size, and output (predicted) field size. One data set was extracted for validation while the rest were used for model training. The model was evaluated based on model output and the validation output data. This was repeated until all data sets had been used for validation purposes. The evaluation data was thereafter analyzed.

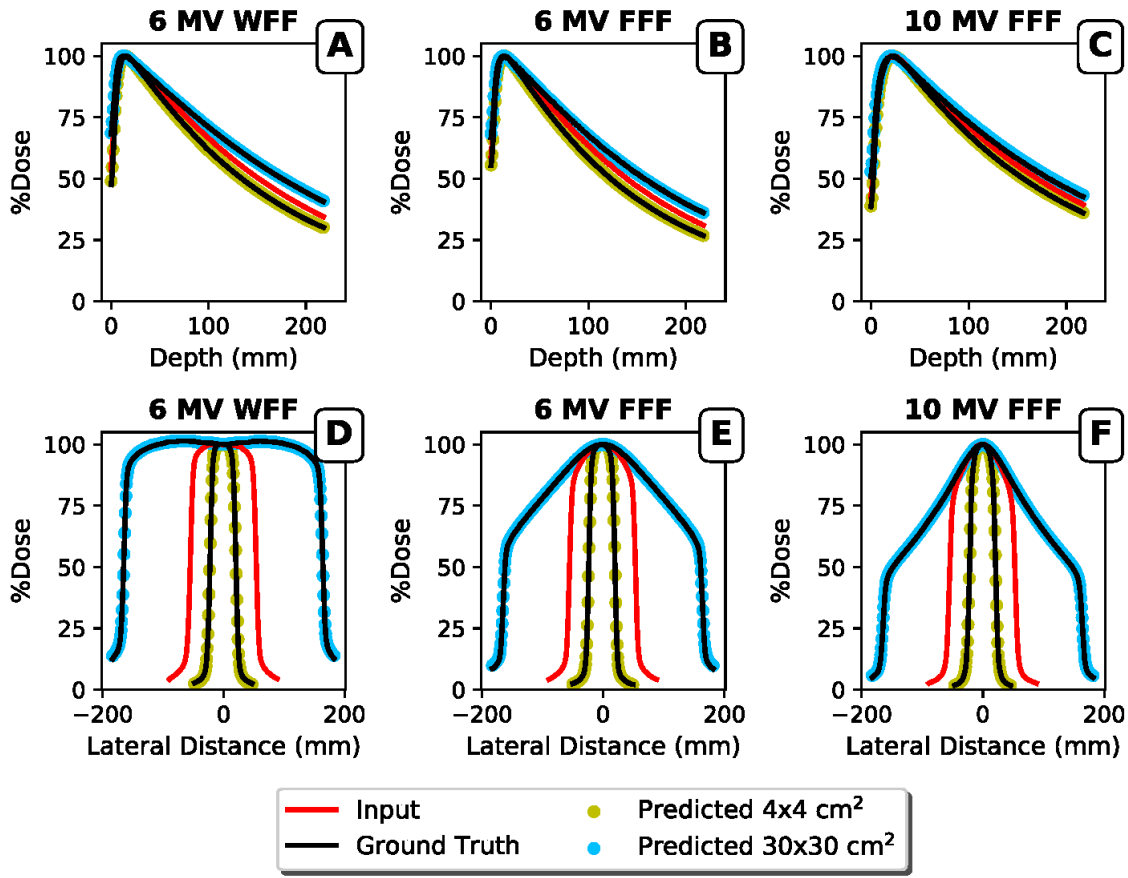

Figure 2. Example of PDD (A, B, C) and profile (D, E, F) prediction of a 4x4 cm$^2$ field (yellow dots) and a 30x30 cm$^2$ field (blue dots) with 10x10 cm$^2$ field as input (red line). Data is shown for the energies 6 MV WFF, 6 MV FFF, and 10 MV FFF. The measured PDDs and profiles (ground truth) for the 4x4 cm$^2$ and 30x30 cm$^2$ beams are depicted as black lines.

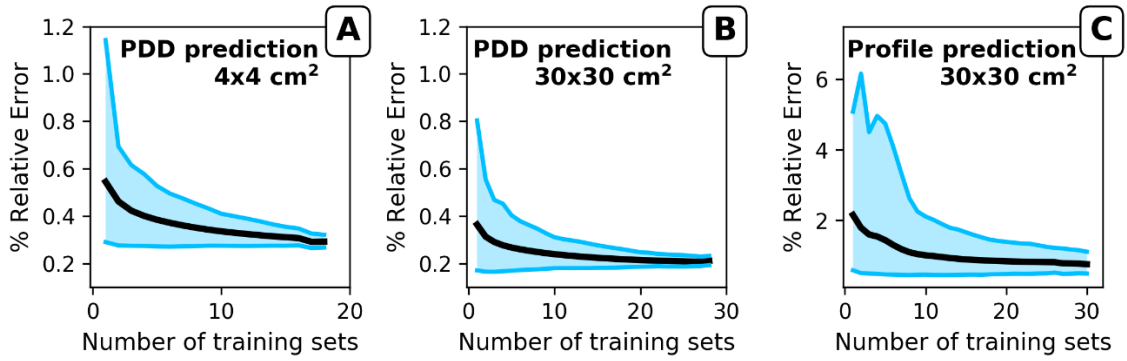

Figure 3. Evaluation of the impact of number of training sets for the prediction of A) PDD for a 4x4 cm$^2$ field, B) PDD for a 30x30 cm$^2$ field, and C) profile for a 30x30 cm$^2$ field. All predictions used a 10x10 cm$^2$ field as input. Graphs show the mean of the absolute %RE (black line) as a function of number of training sets used to build the model. The borders of the blue field represent the maximum and minimum %RE of the predictions.

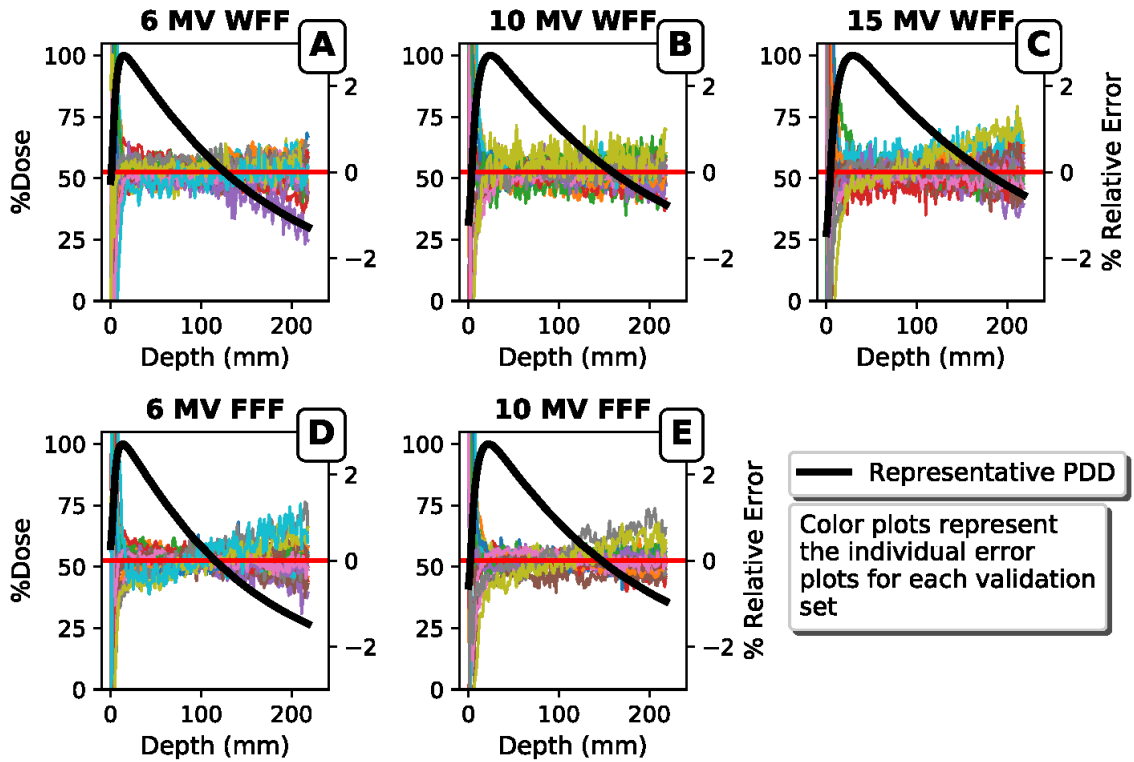

Appendix Figure A.1. Visualization of the 4x4 cm$^2$ field size PDD prediction uncertainty along the entirety of the PDD for each set of data used for validation. The beam PDD for 10x10 cm$^2$ field size was used as input data for the prediction.

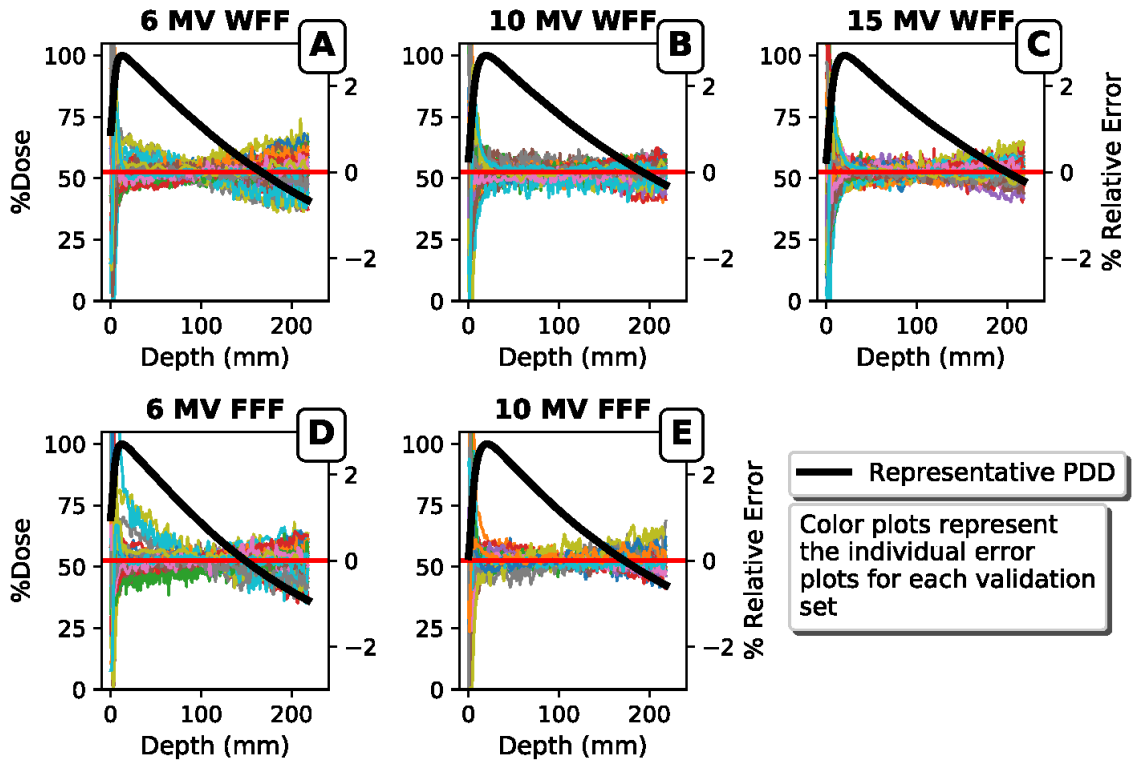

Appendix Figure A.2. Visualization of the 30x30 cm² field size PDD prediction uncertainty along the entirety of the PDD for each set of data used for validation. The beam PDD for 10x10 cm² field size was used as input data for the prediction.

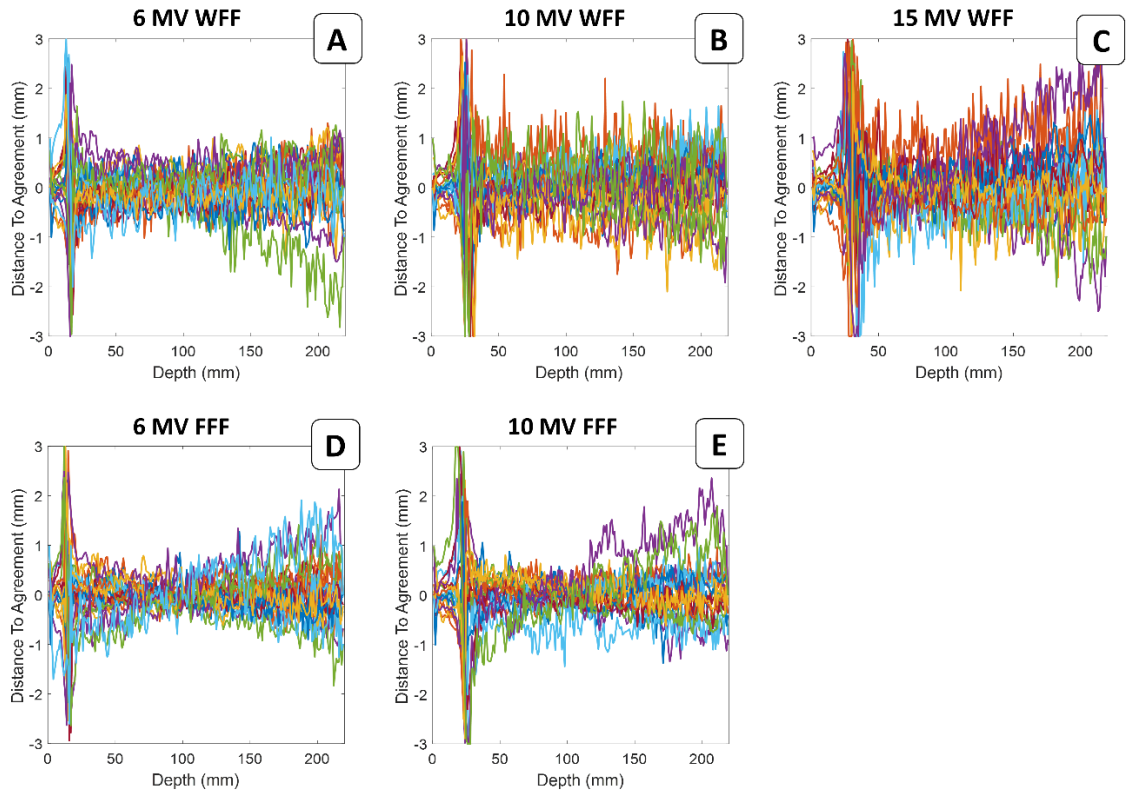

Appendix Figure A.3. DTA distributions for the 4x4 cm$^2$ field size PDD at different beam delivery settings between machine learning predictions and measurements. Line profiles in different colors represent the PDDs for each independent validation set.

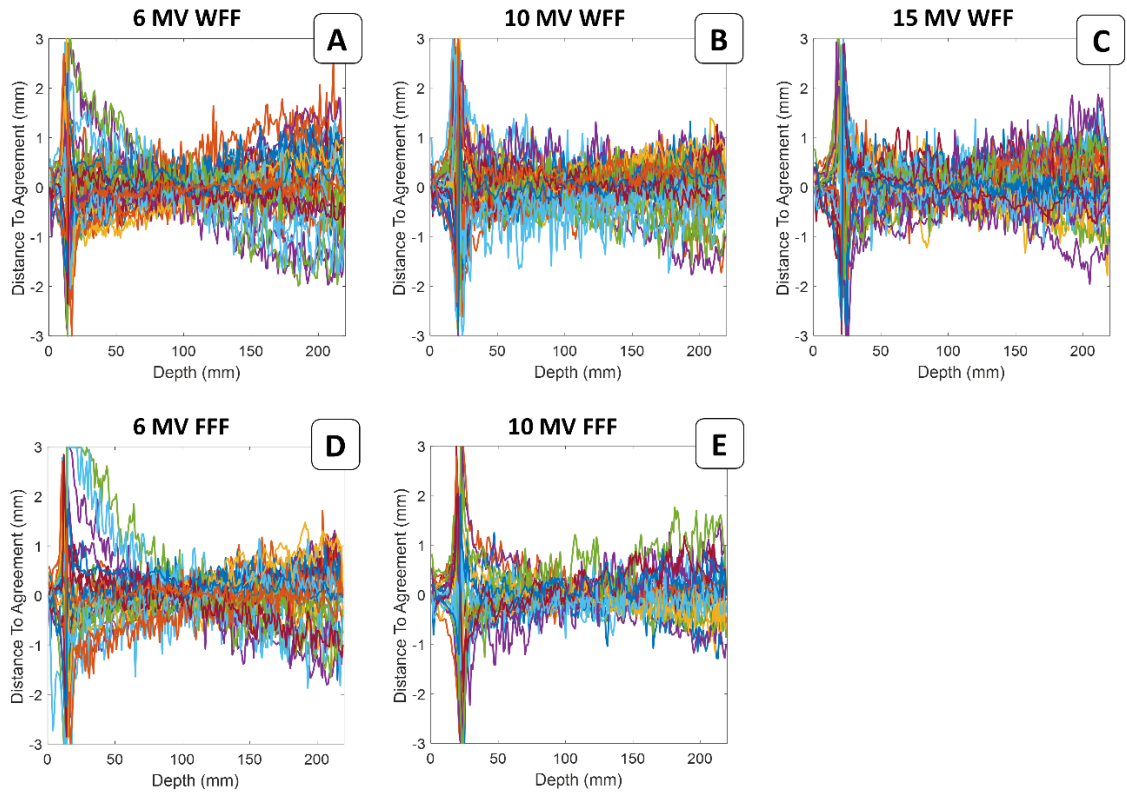

Appendix Figure A.4. DTA distributions for the 30x30 cm$^2$ field size PDD at different beam delivery settings between machine learning predictions and measurements. Line profiles in different colors represent the PDDs for each independent validation set.

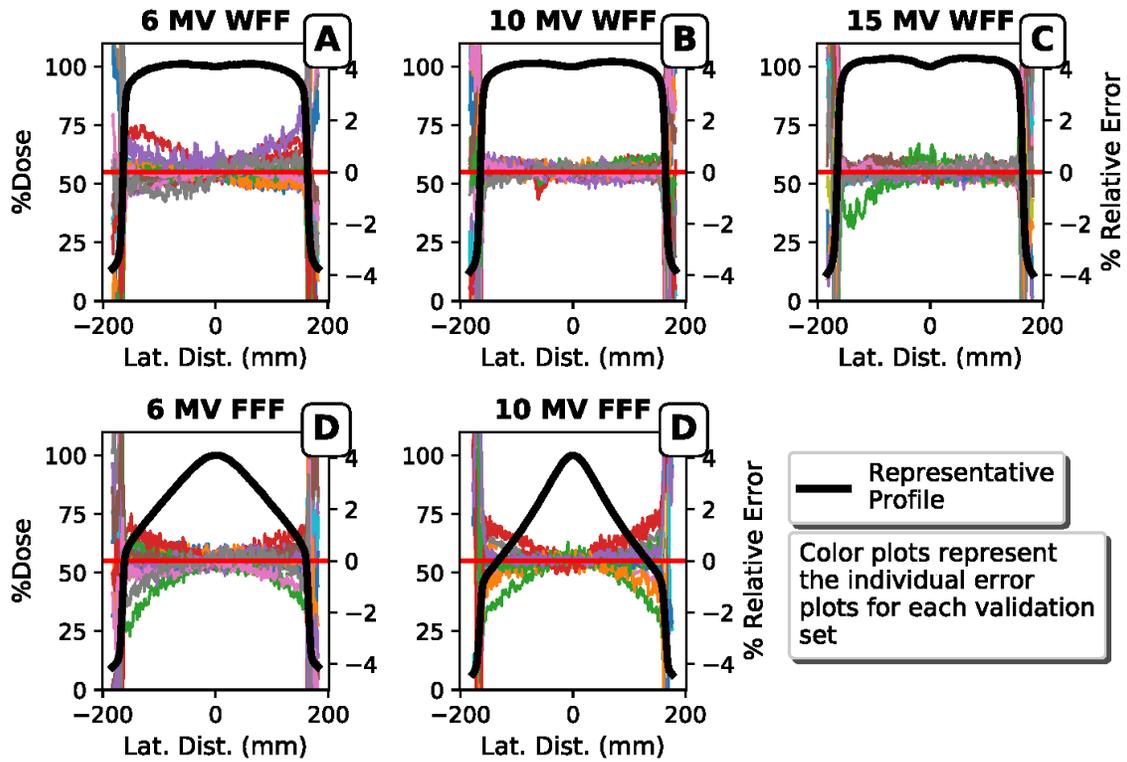

Appendix Figure A.5. Visualization of the 30x30 cm² field size profile prediction uncertainty along the entirety of the profile for each set of data used for validation. The beam profile for 10x10 cm² field size was used as input data for the prediction.

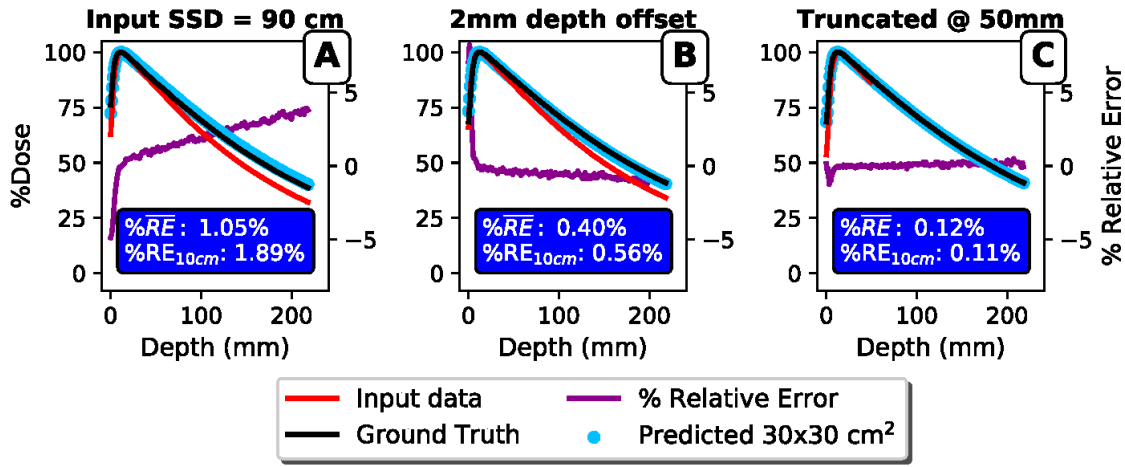

Appendix Figure A.6. Sensitivity of the model to incorrect data input: A) Input data acquired with SSD of 90 cm for model trained on data acquired with SSD of 100 cm, B) input data with a 2mm depth offset, and C) input data truncated at 50 mm depth. The model was trained using the 6 MV WFF beam with input and output data of 10x10 cm² and 30x30 cm², respectively, at SSD 100 cm.

# Tables

Table 1. The absolute percent relative error (%RE) for the predicted PDDs and profiles as well as the absolute %RE for the model input data as compared to the vendor supplied averaged data sets from Varian for TrueBeam linacs. Also shown is the number of data sets used.

| Energy | Field Size | PDD | | | | | Profile | | | | |
|---|---|---|---|---|---|---|---|---|---|---|---|
| | | # of datasets | Model Predicted Beam Data | | Varian Representative Beam Data | | # of datasets | Model Predicted Beam Data | | Varian Representative Beam Data | |
| | | | Mean* | Max | Mean* | Max | | Mean* | Max | Mean* | Max |
| 6 MV WFF | 4 x 4 cm$^2$ | 20 | 0.31 ± 0.10 | 0.51 | 0.42 ± 0.14 | 0.79 | | | | | |
| | 30 x 30 cm$^2$ | 30 | 0.23 ± 0.11 | 0.56 | 0.29 ± 0.15 | 0.74 | 32 | 0.87 ± 0.79 | 3.76 | 1.31 ± 0.68 | 3.81 |
| 10 MV WFF | 4 x 4 cm$^2$ | 19 | 0.31 ± 0.12 | 0.53 | 0.54 ± 0.24 | 1.13 | | | | | |
| | 30 x 30 cm$^2$ | 30 | 0.19 ± 0.06 | 0.37 | 0.34 ± 0.22 | 1.33 | 33 | 0.66 ± 0.38 | 1.68 | 1.36 ± 0.78 | 3.22 |
| 15 MV WFF | 4 x 4 cm$^2$ | 18 | 0.35 ± 0.17 | 0.93 | 0.60 ± 0.33 | 1.34 | | | | | |
| | 30 x 30 cm$^2$ | 29 | 0.19 ± 0.05 | 0.29 | 0.34 ± 0.20 | 1.14 | 31 | 0.71 ± 0.36 | 1.60 | 1.31 ± 0.79 | 3.17 |
| 6 MV FFF | 4 x 4 cm$^2$ | 20 | 0.27 ± 0.16 | 0.77 | 0.35 ± 0.18 | 0.97 | | | | | |
| | 30 x 30 cm$^2$ | 30 | 0.24 ± 0.14 | 0.77 | 0.32 ± 0.22 | 1.19 | 30 | 0.79 ± 0.52 | 2.25 | 1.56 ± 0.89 | 4.97 |
| 10 MV FFF | 4 x 4 cm$^2$ | 19 | 0.26 ± 0.13 | 0.62 | 0.45 ± 0.22 | 1.02 | | | | | |
| | 30 x 30 cm$^2$ | 22 | 0.19 ± 0.07 | 0.40 | 0.54 ± 1.07 | 5.50 | 27 | 0.93 ± 0.45 | 1.79 | 1.80 ± 1.25 | 4.32 |

Note

PDD = Percentage depth dose, WFF = With flattening filter, FFF = Flattening filter free.

* Data are shown as means± standard deviations.

# Tables

Table 2. The absolute percent relative error (%RE) for the predicted PDDs for all different combination of input and output data combinations.

| Energy | Input Field Size | Output field size | Absolute %RE Mean* | Max |
|---|---|---|---|---|
| 6 MV WFF | 4x4 | 10x10 | 0.26 ± 0.08 | 0.46 |
|  | 10x10 | 4x4 | 0.31 ± 0.10 | 0.51 |
|  | 4x4 | 30x30 | 0.29 ± 0.13 | 0.58 |
|  | 30x30 | 4x4 | 0.31 ± 0.09 | 0.52 |
|  | 10x10 | 30x30 | 0.23 ± 0.11 | 0.56 |
|  | 30x30 | 10x10 | 0.24 ± 0.06 | 0.35 |
| 10 MV WFF | 4x4 | 10x10 | 0.24 ± 0.08 | 0.41 |
|  | 10x10 | 4x4 | 0.31 ± 0.12 | 0.53 |
|  | 4x4 | 30x30 | 0.29 ± 0.22 | 1.15 |
|  | 30x30 | 4x4 | 0.36 ± 0.18 | 0.82 |
|  | 10x10 | 30x30 | 0.19 ± 0.06 | 0.37 |
|  | 30x30 | 10x10 | 0.25 ± 0.10 | 0.51 |
| 15 MV WFF | 4x4 | 10x10 | 0.29 ± 0.14 | 0.70 |
|  | 10x10 | 4x4 | 0.35 ± 0.17 | 0.93 |
|  | 4x4 | 30x30 | 0.31 ± 0.25 | 1.21 |
|  | 30x30 | 4x4 | 0.41 ± 0.23 | 1.04 |
|  | 10x10 | 30x30 | 0.19 ± 0.05 | 0.29 |
|  | 30x30 | 10x10 | 0.28 ± 0.16 | 0.75 |
| 6 MV FFF | 4x4 | 10x10 | 0.32 ± 0.20 | 0.96 |
|  | 10x10 | 4x4 | 0.27 ± 0.16 | 0.77 |
|  | 4x4 | 30x30 | 0.37 ± 0.29 | 1.21 |
|  | 30x30 | 4x4 | 0.27 ± 0.14 | 0.71 |
|  | 10x10 | 30x30 | 0.24 ± 0.14 | 0.77 |
|  | 30x30 | 10x10 | 0.29 ± 0.17 | 0.92 |
| 10 MV FFF | 4x4 | 10x10 | 0.22 ± 0.12 | 0.56 |
|  | 10x10 | 4x4 | 0.26 ± 0.13 | 0.62 |
|  | 4x4 | 30x30 | 0.25 ± 0.20 | 0.98 |
|  | 30x30 | 4x4 | 0.29 ± 0.17 | 0.72 |
|  | 10x10 | 30x30 | 0.19 ± 0.07 | 0.40 |
|  | 30x30 | 10x10 | 0.23 ± 0.14 | 0.71 |

Note: WFF = With flattening filter, FFF = Flattening filter free.

* Data are shown as means± standard deviations.

# Tables

Table 3. The absolute percent relative error (%RE) for the predicted profiles for all different combination of input and output data combinations. The mean absolute %RE shows for both the whole profile as well as within the central 80% of the profile based on the size of the predicted field.

| Energy | Input Field Size | Output field size | Absolute %RE | | | |
|---|---|---|---|---|---|---|
| | | | Full profile | | 80% of field size | |
| | | | Mean* | Max | Mean* | Max |
| 6 MV WFF | 10x10 | 30x30 | 0.87 ± 0.79 | 3.76 | 0.14 ± 0.16 | 0.61 |
| | 30x30 | 10x10 | 1.33 ± 1.11 | 4.90 | 0.09 ± 0.05 | 0.18 |
| 10 MV WFF | 10x10 | 30x30 | 0.66 ± 0.38 | 1.68 | 0.08 ± 0.04 | 0.16 |
| | 30x30 | 10x10 | 1.03 ± 0.81 | 4.05 | 0.11 ± 0.09 | 0.31 |
| 15 MV WFF | 10x10 | 30x30 | 0.71 ± 0.36 | 1.60 | 0.11 ± 0.08 | 0.35 |
| | 30x30 | 10x10 | 1.22 ± 1.02 | 3.98 | 0.13 ± 0.13 | 0.56 |
| 6 MV FFF | 10x10 | 30x30 | 0.79 ± 0.52 | 2.25 | 0.17 ± 0.14 | 0.43 |
| | 30x30 | 10x10 | 1.20 ± 0.78 | 3.90 | 0.13 ± 0.14 | 0.6 |
| 10 MV FFF | 10x10 | 30x30 | 0.93 ± 0.45 | 1.79 | 0.14 ± 0.14 | 0.52 |
| | 30x30 | 10x10 | 0.82 ± 0.42 | 1.54 | 0.04 ± 0.03 | 0.14 |

Note: WFF = With flattening filter, FFF = Flattening filter free.

* Data are shown as means± standard deviations.

# Tables

Table A.1. Gamma passing (GP) rates and standard deviations using criteria (2%/1mm and 1%/1mm) for the 4×4 cm² and 30×30 cm² field size PDDs at different beam delivery settings.

| GP (%) | Field size 4×4 cm² | | Field size 30×30 cm² | |
|---|---|---|---|---|
| | 2%/1mm | 1%/1mm | 2%/1mm | 1%/1mm |
| 6 MV WFF | 99.75±0.43 | 99.34±1.45 | 99.91±0.19 | 99.85±0.22 |
| 6 MV FFF | 99.68±0.61 | 99.47±1.14 | 99.77±1.08 | 99.03±3.05 |
| 10 MV WFF | 99.86±0.31 | 99.83±0.31 | 99.94±0.20 | 99.85±0.25 |
| 10 MV FFF | 99.86±0.27 | 99.76±0.44 | 99.94±0.16 | 99.83±0.22 |
| 15 MV WFF | 99.82±0.36 | 99.70±0.86 | 99.92±0.18 | 99.83±0.23 |

Note: WFF = With flattening filter, FFF = Flattening filter free.

* Data are shown as means± standard deviations.